# The mysterious orphans of *Mycoplasmataceae*

Tatiana V. Tatarinova[1,2*], Inna Lysnyansky[3], Yuri V. Nikolsky[4,5,6], and Alexander Bolshoy[7*]

[1] Children's Hospital Los Angeles, Keck School of Medicine, University of Southern California, Los Angeles, 90027, California, USA

[2] Spatial Science Institute, University of Southern California, Los Angeles, 90089, California, USA

[3] Mycoplasma Unit, Division of Avian and Aquatic Diseases, Kimron Veterinary Institute, POB 12, Beit Dagan, 50250, Israel

[4] School of Systems Biology, George Mason University, 10900 University Blvd, MSN 5B3, Manassas, VA 20110, USA

[5] Biomedical Cluster, Skolkovo Foundation, 4 Lugovaya str., Skolkovo Innovation Centre, Mozhajskij region, Moscow, 143026, Russian Federation

[6] Vavilov Institute of General Genetics, Moscow, Russian Federation

[7] Department of Evolutionary and Environmental Biology and Institute of Evolution, University of Haifa, Israel

1,2 *tatiana.tatarinova@usc.edu*

3 *innal@moag.gov.il*

4-6 *YNikolskiy@sk.ru*

7 *bolshoy@research.haifa.ac.il*

## Abstract

**Background**: The length of a protein sequence is largely determined by its function, i.e. each functional group is associated with an optimal size. However, comparative genomics revealed that proteins' length may be affected by additional factors. In 2002 it was shown that in bacterium *Escherichia coli* and the archaeon *Archaeoglobus fulgidus,* protein sequences with no homologs are, on average, shorter than those with homologs [1]. Most experts now agree that the length distributions are distinctly different between protein sequences with and without homologs in bacterial and archaeal genomes. In this study, we examine this postulate by a comprehensive analysis of all annotated prokaryotic genomes and focusing on certain exceptions.

**Results**: We compared lengths' distributions of "having homologs proteins" (HHPs) and "non-having homologs proteins" (orphans or ORFans) in all currently annotated completely sequenced prokaryotic genomes. As expected, the HHPs and ORFans have strikingly different length distributions in almost all genomes. As previously established, the HHPs, indeed, are, on average, longer than the ORFans, and the length distributions for the ORFans have a relatively narrow peak, in contrast to the HHPs, whose lengths spread over a wider range of values. However, about thirty genomes do not obey these rules. Practically all genomes of *Mycoplasma* and *Ureaplasma* have atypical ORFans distributions, with the mean lengths of ORFan larger than the mean lengths of HHPs. These genera constitute over 80% of atypical genomes.

**Conclusions:** We confirmed on a ubiquitous set of genomes the previous observation that HHPs and ORFans have different gene length distributions. We also showed that *Mycoplasmataceae* genomes have very distinctive distributions of ORFans lengths. We offer several possible biological explanations of this phenomenon.

# Background

Different factors affect properties of prokaryotic proteins. Some of them appear to be general constraints on protein evolution. For example, genomic studies revealed that the base composition of a genome (i.e. GC content) correlates with the overall amino acid composition of its proteins [2]. There are also general constraints on protein size, such as, in general, smaller proteins for prokaryotes compare to eukaryotes [3]. Previously, we revealed some other factors affecting the lengths of protein-encoding genes [4-6]. However, there are numerous protein-encoding genes without homologues in genomes of other organisms called "ORFans" or "orphans" (the term coined by Fisher and Eisenberg [7]). The ORFans are not linked by overall similarity or shared domains to the genes or gene families characterized in other organisms. Tautz and Domazet-Lošo [8] were the first to discuss systematic identification of ORFan genes in the context of gene emergence through duplication and rearrangement processes. Their study was supported by other excellent reviews [9-11].

ORFan genes were initially described in yeast as a finding of the yeast genome-sequencing project [12, 13], followed by identification of ORFans in all sequenced bacterial genomes. Comparative genomics has shown that ORFans are an universal feature of any genome, with a fraction of ORFan genes varying between 10-30% per a bacterial genome [14]. Fukuchi and Nishikawa [15] identified that neither organism complexity nor genome length correlate with the percentage of ORFan genes in a genome.

ORFans are defined as the genes sharing no similarity with genes or coding sequence domains in other evolutionary lineages [12, 13]. They have no recognizable homologs in distantly related species. This definition is conceptually simple, but operationally complex. Identification of ORFans depends both on the detection method and the reference set of genomes, as this defines the evolutionary lineage to be investigated. Albà and Castresana [16] have questioned whether BLAST was a suitable procedure to detect all true homologues and concluded that BLAST was a proper algorithm to identify the majority of remote homologues (if they existed). Tautz and Domazet-Lošo developed a general framework, the so-called "Phylostratigraphy" [17], which consists of statistical evaluation of macro-evolutionary trends [17-19]. Phylostratigraphy is being applied for systematic identification of all orphan genes within the evolutionary lineages that have led to a particular extant genome [18-23].

Lipman et al. [1] studied the length distributions of the Having Homologs Proteins (HHP) and Non-Conserved Proteins (ORFans in our nomenclature) sets for the bacterium *Escherichia coli*, the archaeon *Archaeoglobus fulgidus*, and three eukaryotes. Regarding the two prokaryotes, the group made the following observations:

i.  HHPs are, on average, longer than ORFans.
ii. The length distribution of ORFans in a genome has a relatively narrow peak, whereas the HHPs are spread over a wider range of values.

Lipman et al. [1] proposed that there is a significant evolutionary trend favoring shorter proteins in the absence of other, more specific functional constraints. However, so far research in this area was limited in the scope of organisms. Here, we have tested the above-mentioned observations by Lipman et al. [1] on a comprehensive set of all sequenced and annotated bacterial genomes. We performed comparisons of length distributions of HHP and ORFans in all annotated genomes and confirmed, to a large extend, the conclusions of Lipman et al. [1]. Below, we described and discussed the few remarkable exceptions to the general rules.

## Results and discussion

Most exceptions species to the Lipman's rule [1] belong to the *Mycoplasmataceae* family. *Mycoplasmataceae* lack the cell wall, feature some of the smallest genomes known and are "metabolically challenged", i.e missing some essential pathways of free-living organisms [24-28]. Many *Mycoplasmataceae* species are pathogenic in humans and animals.

### HHPs and ORFans lengths

We have selected four genomes out of currently sequenced and annotated 1484 bacterial genomes to illustrate typical protein lengths distributions for HHPs and ORFans, (Fig. 1, Panels A-D). The ORFans' length distributions are relatively narrow, in contrast to the HHPs with lengths spread over a wider range of values. ORFans are obviously shorter than HHPs in all four species (Fig. 1, Panels A-D). Note that the distributions of protein lengths in the four selected bacteria are similar to the global distribution presented in Fig. 1 (Panels E-F).

Based on the data from two genomes, Lipman *et al.* [1] suggested that HHPs are,

on average, longer than the ORFan proteins in general case.

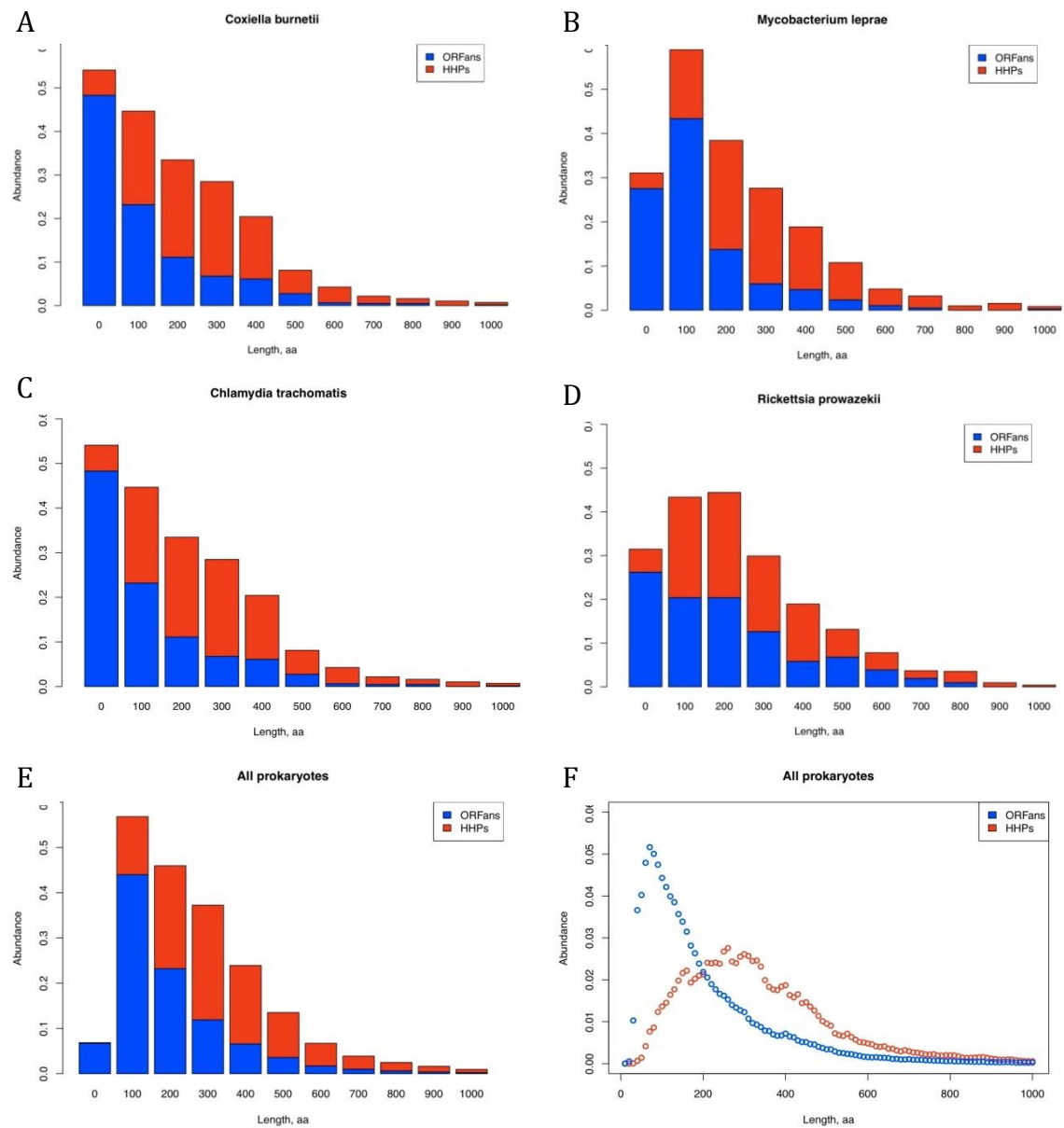

Figure 1: Figure 1. Histograms of protein lengths of *Coxiella burnetii* (A), *Mycobacterium leprae* (B), *Chlamydia trachomatis* (C), *Rickettsia prowazekii* (D) and all other prokaryotes (E and F) tested in this study. Stacked bar plot (E) and the relative frequency plot with a smaller bin size (F). Axis X corresponds to the following protein length intervals 0 -- (0,100], 100-- (100,200], etc. . Stacked bar plot and the relative frequency plot with a smaller bin size are presented on Panels E and F, respectively.

In order to test this statement, we have calculated distributions of protein lengths for all COG-annotated genomes, and built a histogram of differences between the means of HHPs and ORFans, which happened to be approximately bell-shaped (Fig. 2). On average, HHPs are longer than ORFans by approximately 150 amino acids. However, the bell-shaped distribution has a heavy left tail containing genomes with ORFans' mean length equal to or exceeding HHPs' mean length (Figure 2). In order to investigate this effect, we sorted the genomes according to the difference between the mean lengths of HHPs and ORFans (Table 2). Most "atypical genomes" with longer ORFans belong to the species from *Mycoplasmataceae* family (*Mycoplasma* and *Ureaplasma* genera) and some to the *Anaplasmataceae* family (*Anaplasma* and *Ehrlichia* genera). There are also solitary representatives of other lineages: *Chlorobium chlorochromatii, Lawsonia intracellularis, Burkholderia pseudomallei, Rhodobacter sphaeroides,* and *Methanobrevibacter ruminantium.* Only the *Mycoplasmataceae* family contains 32 fully sequenced and annotated genomes with atypical ORFans, which is sufficient for statistical analysis (see Table 1). Therefore, we restricted our analysis to the *Mycoplasma* and *Ureaplasma* genera.

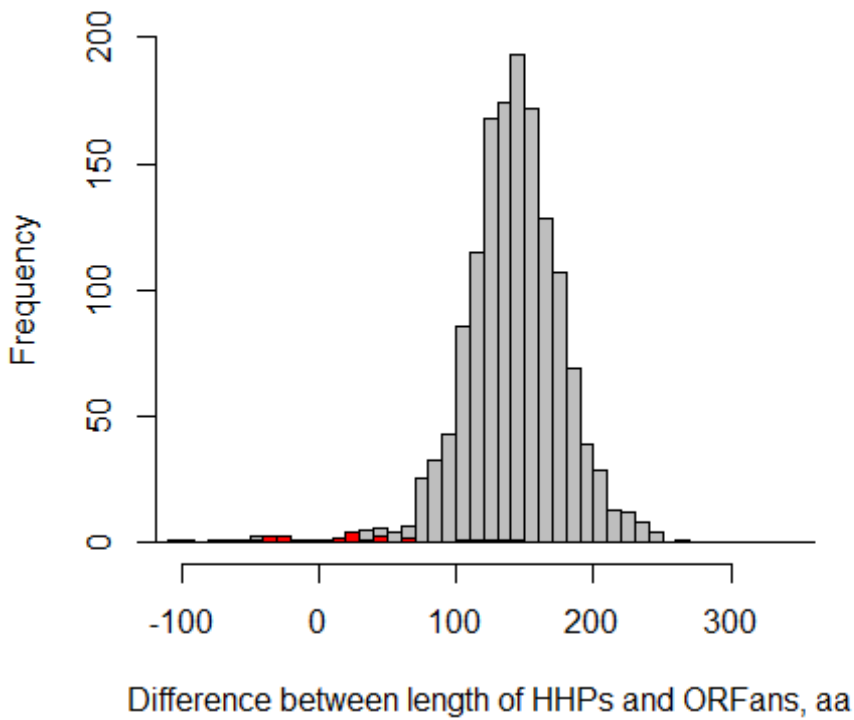

**Figure 2: The difference between mean lengths of HHPs and ORFans for 1484 prokaryotic genomes. Histogram for Mycoplasmataceae genomes is in red and for all other species is in grey.**

**Variability of protein lengths**

The *Mycoplasmataceae* genomes challenge the second conclusion of Lipman *et al.* [1] that the length distributions of ORFans have a relatively narrow peak, whereas those of the HHP are spread over a wider range of values. The histogram of differences between HHPs and ORFans in these atypical genomes is shown in Fig. 2 (red bars). We calculated the Correlation of Variation ( $CV = \frac{sd(Y)}{\bar{Y}}$ , where Y is a set of protein lengths); average difference between CV for ORFans and HHPs in "atypical" genomes was 0.31. We also computed variances of lengths for ORFans and HHPs separately and conducted the F-test, resulting in p-values <10$^{-64}$ for all tested pairs. Therefore, the ORFan proteins of these genomes are more variable in length than the HHPs.

**Selection of a statistic for identification of atypical genomes**

We tested the relationships between the mean HHP length and the mean ORFan length on eight groups of prokaryotes: two families of *Mycoplasmataceae* and *Mycobacteriaceae*, six genera *Agrobacterium, Bacillus, Anaplasma, Ehrlichia, Neorickettsia* and *Campylobacter* (Fig. 3, Panel A). *Mycoplasmataceae* genomes form a clearly distinct group of atypical genomes. As shown below, there is a small number of unusually long ORFan proteins in *Mycoplasmataceae*, the outliers that may skew the distribution. Therefore, considering only the mean gene lengths distribution may be insufficient; the median value is, probably, a more appropriate measure (Fig. 3, Panel B). However, *Mycoplasmataceae*, again, represent a group of atypical genomes. Therefore, the effect cannot be explained exclusively by poorly predicted outliers in the *Mycoplasmataceae* genomes.

A                                    B

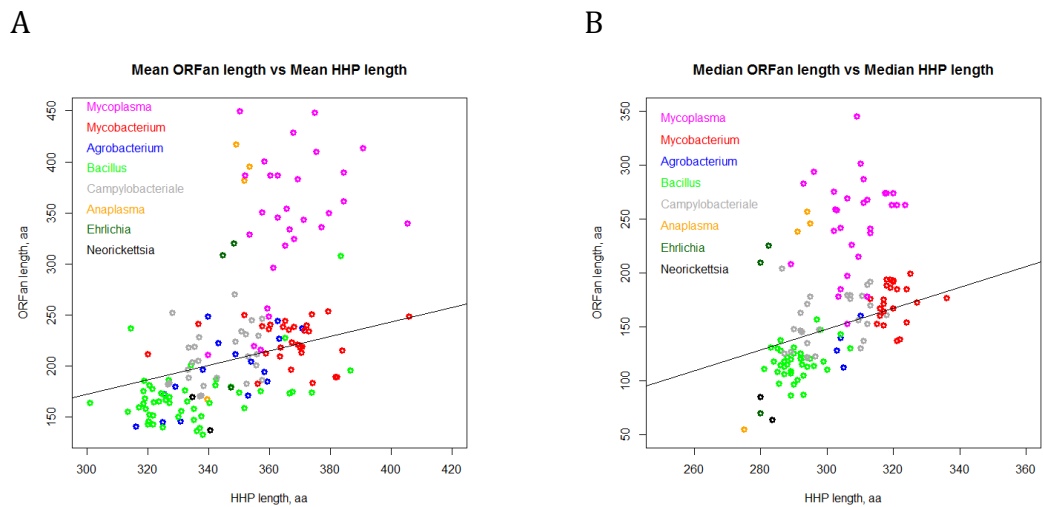

**Figure 3: Mean (A) and Median (B) ORFans' length vs. average HPP length for selected eight groups of prokaryotes. Family Mycoplasmataceae (pink), family Mycobacteriaceae (red), genus Agrobacterium (blue), genus Bacillus (green), genus Anaplasma (orange), genus Ehrlichia (dark green), genus Neorickettsia (black) and genus Campylobacter (grey) are shown. The regression line shows the relationships between the mean HHP length and the mean ORFan length across 1484 annotated prokaryotic genomes.**

It is worth mentioning that outside of *Mycoplasma* and *Ureaplasma* genera, there is one currently sequenced bacterial genome with ORFans' mean length larger than the HHPs. In *Candidatus Blochmannia floridanus*, the former value is twice as large as the mean length of HHPs, due to the only unusually large ORFan protein of 680 amino acids (while the average HHP length for *Candidatus Blochmannia floridanus* is 334 aa, median length is 294 aa and the longest protein is 1420 aa).

Typical distributions of protein length of HHPs and ORFans in *Mycoplasmataceae* are illustrated by two genomes (*M. genitalium* and *M. hyopneumonia*) selected out of 68 sequenced genomes of *Mycoplasmataceae* (Fig.4).

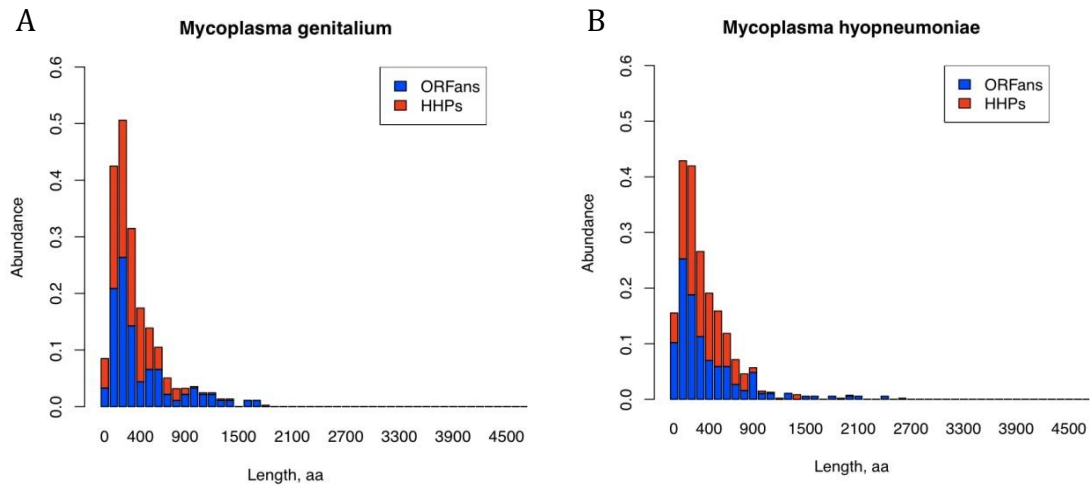

**Figure 4: Histograms of protein lengths of *M. genitalium* and *M. hyopneumonia*. X axis labels correspond to the following protein length intervals 0 -- (0,100], 100-- (100,200], etc. Y axis shows a relative frequency of the protein with a given length in a genome.**

These ORFans' distributions are rather different from the ones for four bacteria shown above (Fig. 1, Panels A-D). *Mycoplasma* protein lengths distributions have two properties that distinguish them from other organisms:

i. Existence/presence of few very short ORFans in Mycoplasma' genomes. On average, *Mycoplasma* species contain 44±12 short (length <100 aa) proteins per genome as compared to 145±7 short proteins per genome in all other sequenced prokaryotes.

ii. Comparatively many ORFans longer than 800 aa (on average, *Mycoplasma* species contain 17±4 long (≥800 aa) proteins per genome as compared to 11±1 long proteins per genome in all other sequenced prokaryotes). Moreover, there are several extremely long ORFans. On average, *Mycoplasma* species contain 10±1 very long (≥1000 aa) proteins per genome as compared to 6±1 very long proteins per genome for all other sequenced prokaryotes.

**Functional annotation of ORFans**

We selected a 9350 ORFans of from 32 species *Mycoplasmataceae*, found the best hits in other prokaryotic genomes, and stratified them by functional annotation in the COG database. 54% of ORFans were mapped to a "hypothetical protein" category; 6% are 'lipoproteins", further 2% are "membrane lipoproteins"; 3% are "surface protein 26-residue repeat-containing proteins", and the rest is mapped to lesser-abundant categories. A protein is called "hypothetical" if its existence has been predicted *in silico*, but the function is not experimentally validated. Despite *Mycoplasmataceae* cells are wall-less with no periplasmic space, they effectively anchor and expose surface antigens using acylated proteins with long-chain fatty acids [29-31]. Lipoproteins are abundant in mycoplasmal membranes and are considered to be a key element for diversification the antigenic character of the mycoplasmal cell surface [29, 32].

For the long proteins (≥1000 aa) we are especially interested in, we compared the functional annotations between HHPs and ORFans. These two groups were most different in the "hypothetical protein" category (p-value=4.00195E-21) overrepresented in ORFans, followed by "efflux ABC transporter, permease protein", also over-represented in the long ORFans of *Mycoplasmataceae* (p-value =4.18953E-06). The best BLAST hits of *Mycoplasma*'s "efflux ABC transporter, permease proteins" were to the ABC transporter proteins from two related species, *Ureaplasma parvum* and *Ureaplasma urealyticum.* Moreover, multiple protein alignment in CLUSTALW (see Supplementary Data) shown high degree of conservation among the "efflux ABC transporter, permease proteins" across all genomes of *Mycoplasmataceae*.

**Are the observed peculiarities features of *Mycoplasmataceae* family or the entire class of *Mollicutes*?**

In order to investigate whether long ORFans is a specific feature of *Mycoplasmacaea,* we analyzed ORFans sizes in several species from the same *Mollicutes* class, including *Acholeplasma* and *Candidatus Phytoplasma*. In *Candidatus Phytoplasma australiense* and *Acholeplasma laidlawii*, ORFans are 1.5-2 times shorter than HHPs. Therefore, we concluded that these features are not universal for *Mollicutes*.

We have also analyzed the genomes of *Anaplasma* that belong to a family *Ehrlichiaceae* in the order of *Rickettsiales*. The genus *Anaplasma* includes obligatory parasitic intracellular bacteria, residing in the vacuoles in eukaryotic host cells and lacking stained cytoplasm. On average, in *Anaplasma centrale*, the ORFan proteins are 68 aa longer than of HHP proteins, and in *Anaplasma marginale* ORFans are 42 aa longer than HHPs. However, the median protein lengths of ORFans are 40 and 50 aa shorter than HHPs in both considered *Anaplasma* genomes, correspondingly (see Figure 3). This discrepancy is due to several unusually long ORFan proteins with hypothetical function that skew the mean length up. Moreover, the ORFans feature shorter mean and median lengths than HHPs in all tested *Ehrlichia* and *Neorickettsia* species (*Ehrlichia canis, Ehrlichia ruminantium, Ehrlichia chaffeensis, Neorickettsia risticii, Neorickettsia sennetsu*). These bacteria (together with *Anaplasma* species) belong to the order *Rickettsiales*. Two strains of *Ehrlichia ruminantium Welgevonden* were excluded due to an inconsistency of annotation between them. Based on the data obtained, we concluded that the phenomenon of extremely long ORFans is specific for the family of *Mycoplasmataceae*.

**Driving forces behind the long ORFans**

Why the *Mycoplasmataceae* have ORFans as long as HHPs with the distribution of ORFans' lengths very similar to HHPs? *Mycoplasmataceae* are a heterogeneous group of the cell-wall-less, the smallest and the simplest self-replicating prokaryotes. They have a reduced coding capacity and have lost many metabolic pathways, as a result of parasitic lifestyle [33, 34]. These organisms are characterized by lack of a cell wall, small genome size, low G+C content (23% to 40%) and atypical genetic code usage (UGA encodes tryptophan instead of a canonical opal stop codon) [35]. In addition, *Mycoplasmataceae* genomes lack of 5' UTR in mRNAs as established by Nakagawa et al. [36]. This phenomenon is highly unusual in bacteria. Below we propose and discuss several reasons that might explain the presence of long ORFans in *Mycoplasmataceae*.

 1. **Small genome size**

Prokaryotic genomes range from 10Kbp (*Bacteroides uniformis*, associated with the degradation of the isoflavone genistein in human feces) to 39 Mbp (*Vibrio parahaemolyticus*, causing acute gastroenteritis in humans), with the mean length of 3.5 Mbp and the median of 3.0 Mbp [37]. *Mycoplasmataceae*, indeed, tend to have small genomes (mean/median lengths are 0.9 Mbp, minimum is 0.58 Mbp, and the maximum is 1.4 Mbp). However, there are many bacteria with smaller genomes, including such "dwarfs" as *Candidatus Tremblaya princeps* and *Candidatus Hodgkinia cicadicola* (0.14 Mbp each), and *Candidatus Carsonella ruddii* (0.17 Mbp). The "genomic dwarfism" per se is not associated with unusual ORFans. Among the 324 annotated "genomic dwarfs" with genome sizes below 2 Mbp, only *Ureaplasma, Anaplasma* and *Mycoplasma* genomes feature the average ORFan length to be over 95% of the average HHP length. In all other species (except one), the ratio of ORFan to HHP length ranges from 30% to 90%. The exception is a tiny (400 nm in diameter) marine Archaeon *Nanoarchaeum equitans* with the average ORFans' length of 94% of the HHPs' length. N*anoarchaeum* is a remarkable organism; it is an obligate symbiont on the archaeon *Ignicoccu*s, which cannot synthesize lipids, amino acids, nucleotides, or cofactors [38].

Neither β-proteobacterium *Candidatus Tremblaya princeps* (endosymbiotic bacteria living inside the citrus pest mealybug *Planococcus citri*), *Candidatus Carsonella ruddii* (endosymbiont present sap-feeding insects psyllids), nor *Candidatus Hodgkinia cicadicola* (α-proteobacterial symbiont of cicadas) feature unusually long ORFans. For all three species, the mean ORFan length is approximately 40% of the HHP length. Therefore, we conclude that the small genome size alone cannot explain the presence of long ORFans in *Mycoplasmataceae*.

## 2. Low GC content and unusual base composition in a reduced bacterial genome

We analyzed 300 genomes with the lowest GC content (ranging from 14% to 36%), including 3 species of *Ureaplasma* and 27 species of *Mycoplasma*. Overall, there is only a weak positive correlation (Pearson's correlation coefficient $\rho = 0.13$) between the GC-content and the ORFan to HHP length ratio, and plenty of examples of GC-poor genomes with low ORFan to HHP length ratio. The GC-poor species feature an average ORFans to HHPs ratio of 60%, ranging from 20% to 106%. Among 10 most GC-poor genomes, the ORFan to HHP length ratio varies between 30% and 76%.

The GC-poor *Ureaplasma* and *Mycoplasma* species have average ORFans to HHPs ratio of 98%, the lowest being 61% and the highest 130%. Interestingly, among the 300 GC-poor species the upper tail of the high ORFan to HHP length ratio is occupied by *Ehrlichia ruminantium, Ehrlichia canis, Methanobrevibacter ruminantium,* and *Nanoarchaeum equitans*. *E. ruminantium* and *E. canis* belong to the *Anaplasmataceae* family; they are obligatory intracellular pathogens transmitted by ticks. According to Mavromatis et al. [39], *E. canis* genome contains a large number of proteins with transmembrane helices and/or signal sequences and a unique serine-threonine bias prominent in proteins associated with pathogen-host interactions.

The $GC_3$ is defined as a fraction of guanines and cytosines in the third codon position [40]. The importance of variability in genomic GC and genic $GC_3$ content for stress adaptation has been established by multiple authors for a number of prokaryotic and eukaryotic organisms [41-45]. The mechanisms behind GC-content differences in bacterial genomes are unclear, although variability in the replication and/or repair pathways were suggested as hypotheses [46-48]. One mechanistic clue is the positive correlation between the genome size and GC content (smaller genomes tend to have lower GC-content). This tendency is particularly pronounced for obligate intracellular parasites. Two (not necessarily mutually exclusive) hypotheses have been forwarded to explain this base composition bias in the genomes of intracellular organisms. The first is an adaptive hypothesis, based on selection for energy constraints [49]. It stays that low GC content helps the intracellular parasites to compete with the host pathways for the limited metabolic resources in cytoplasm. The second hypothesis relates to mutational pressure resulting from the limited DNA repair systems in bacterial parasites [50]. Small intracellular bacteria often lose non-essential repair genes, and, therefore, are expected to be deficient in their ability to repair damage caused by spontaneous chemical changes. This is particularly expected for endosymbionts, in which genetic drift plays a major role in sequence evolution [50].

Thus, *Mycoplasma,* and *Ureaplasmae* are GC and $GC_3$ - poor, (Figure 5, Supplemental Table 1). Why GC-poverty is so important? According to the "codon capture model", in GC–poor environment, the replication mutational bias towards AT causes the stop codon TGA to change to the stop codon TAA without affecting protein length [51, 52]. The subsequent appearance of the TGA codon through a point mutation leaves it free to encode for an amino acid (Trp). This brings us to our next point of discussion.

A                                                    B

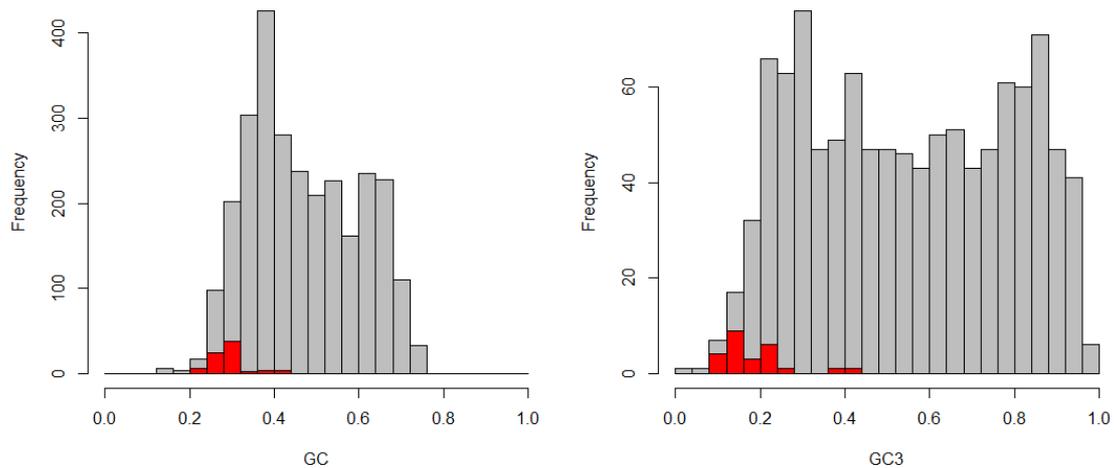

**Figure 5:** Genomic GC content (Panel A) and genic GC3 content (Panel B) in annotated species of Mycoplasmataceae. Grey histograms correspond to all prokaryotes while red histograms correspond to selected Mycoplasma species.

In *Mycoplasmataceae,* ORFans have 3% lower GC content than HHPs. This is close to the average difference in GC content for ORFan genes in all prokaryotic species (-3.9%, with the range from -25% to 25%). Some species with the lowest GC-content in ORFans are *Dickeya dadantii, Citrobacter rodentium, Pectobacterium wasabiae, Chromobacterium violaceum, Alicyclobacillus acidocaldarius, Neisseria meningitidis,* and *Shigella sonnei.* The species with the highest GC content in ORFans include *Methylobacterium chloromethanicum CM4, Escherichia coli SE11, Lactobacillus plantarum, Spirosoma linguale* and *Bifidobacterium longum infantis.*

Variability of $GC_3$ content in bacteria appears to be an instrument of environmental adaptation, allowing keeping the protein sequence unchanged. According to Mann and Chen [53], in nutrient limiting and nutrient poor environments, smaller genome size and lower GC content help to conserve replication expense. Generally, species with many $GC_3$-rich genes have ORFans with lower $GC_3$ content, and species with many $GC_3$-poor genes (average $GC_3<0.3$) have ORFans with the same or higher $GC_3$ content as HHPs. We observed that, on average, prokaryotic ORFans have 12.5% lower $GC_3$ content as compared to HHPs of the same organism. Some species (such as *Burkholderia pseudomallei, Burkholderia mallei, Thermobispora bispora, Burkholderia pseudomallei, Chromobacterium violaceum, Rhodobacter sphaeroides,* and *Kineococcus radiotolerans*) feature two-fold decrease in $GC_3$ content of ORFans as compared to HHPs. ORFans of some other species have higher $GC_3$ content than HHPs (20% increase or more). These include

*Methylobacterium chloromethanicum, Pelagibacterium halotolerans, Escherichia coli SE11, Lactobacillus plantarum,* and *Thermofilum pendens.* Curiously, ORFans of *Mycoplasma* and *Ureaplasma* have the same $GC_3$ content as their HHPs (around 20%). It appears that, since the genes of *Mycoplasma* and *Ureaplasma* already have low $GC_3$ content, there simply is no room to decrease it further it for ORFans.

Based on these findings, we conclude that GC-content of genes and genome cannot be a sole factor responsible for existence of long ORFans in a *Mycoplasmataceae*.

### 3. UGA StopRTrp recoding

Almost all bacterial and archaeal species have three stop codons: TAA, TGA and TAG. However, there are 77 exceptions to this rule among the currently completely sequenced 2723 prokaryotic genomes (note that only 1484 of them are COG-annotated and, therefore, were used in our study). Seventy-three species out of seventy-seven belong to the genera *Mycoplasma, Spiroplasma*, and *Ureaplasma*; all of them are small bacteria of the class *Mollicutes*. In addition, in several mitochondrial lineages, the UGA StopRTrp recoding is also associated with both genome reduction and low GC content [54-56]. For example, *Candidatus Hodgkinia cicadicola*, mentioned above because of its "dwarf genome", also features the coding reassignment of UGA Stop→Trp [57]. Moreover, two groups of currently uncultivable bacteria, found in marine and fresh-water environment and in the intestines and oral cavities of mammals, use UGA as an additional glycine codon instead of a signal for translation termination [58]. Under the "codon capture" model, a codon falls to low frequency and is then free to be reassigned without major fitness repercussions. Applying this model to the UGA StopRTrp recoding, mutational bias towards AT causes each UGA to mutate to the synonym UAA without affecting protein length [51, 52]. When the UGA codon subsequently reappears through a mutation, it is then free to encode for an amino acid [51, 52]. While some have argued that codon capture is insufficient to explain many recoding events [2, 54, 55], the fact that all known UGA StopRTrp recoding has taken place in low GC genomes [54, 59] makes the argument attractive for this recoding. It was suggested [51] that the recoding is driven by the loss of translational release factor RF2, which recognizes the TGA stop codon. Notably, despite the fact that *Candidatus Hodgkinia cicadicola* uses the UGA StopRTrp recoding, it has perfectly normal difference of distributions between ORFans and HHPs [5, 6]. According to Ivanova et al. [60], TGA reassignment is likely to be limited to the *Mollucites* and *Candidatus Hodgkinia cicadicola,* and it occurred as a single event after the last common ancestor separated from the *Peregrine*s group.

We have also examined other members of *Mollicutes* and found numerous examples when the distribution of differences between ORFans and HHPs was normal. Since 73 out of 77 species with TGA reassignment are *Mycoplasma* and *Ureaplasma* species, we do not have enough statistical power/data to conclude whether recoding of UGA StopRTrp is the main cause of long ORFans.

4. **Lack of a cell wall and parasitic lifestyle**

Several bacterial species have wall-less cells (L-forms), as a response to extreme nutritional conditions [61]; L-forms might have played a role in evolution with respect to the emergence of *Mycoplasma* [62]. In order to compensate for the lack of cell wall, *Mycoplasma* developed extremely tough membranes capable to contend with the host cell factors. Lipoproteins are abundant in mycoplasmal membranes [29, 32]. They modulate the host's immune system [63], therefore playing an important role in the infection propagation. Ability of lipoproteins to undergo frequent size or phase variation is considered to be an adaptation to different conditions, including the host's immune response [63, 64]. Some of the largest gene families in *Mollicute* genomes encode ABC transporters, lipoproteins, adhesins and other secreted virulence factors [31]. This may be due to the absence of a cell wall and a periplasmic space in *Mollicute*, attributable to their parasitic lifestyle in a wide range of hosts. We identified many of the *Mollicutes* ORFans as hypothetical proteins and lipoproteins in the COG functional classification. Moreover, hypothetical proteins and efflux ABC transporter, permease proteins were predominant among the longest proteins (≥1000 aa). Hypothetical proteins constitute a large group proteins in *Mollicutes* [65-67]. Lipoproteins, especially membrane exposed ones, are abundant in *Mollicutes*, in sharp contrast to other bacteria, which only have a limited number of lipoproteins in the membranes [31]. In general, lipoproteins carry out numerous important functions, including protection against osmotic and mechanical stress and interactions with the host [31]. However, most *Mollicutes* lipoproteins currently lack the exact functions and their host protein interaction partners are unknown. Depending on the species, lipoproteins are encoded by a single or multiple genes (multi-gene families) and some of them are members of paralogous families, such as P35 lipoprotein of *M. penetrans* [68]. Some lipoproteins are species-specific, while some have homologs among different species, in particular, are associated or share sequence similarity with ABC transporter genes, suggesting that they may play a role in the transport of nutrients into the cell [69]. It is well established that prokaryotic ABC transporters translocate different compounds across cellular membranes in an ATP coupled process (a crucial function for obligate parasites like *Mollicutes*). They also carry out a remarkable diversity of other functions, some of which are essential for pathogenicity [70].

The accessory genes or ORFans are usually an important source of genetic variability in bacterial populations, which thought to play a role in niche adaptation, host specificity, virulence or antibiotic resistance. Most of the identified *Mycoplasmataceae* ORFans are surface exposed proteins, which suggests that they may play a role in shielding the wall-less mycoplasma cell membrane from host defense. Interestingly, the long variable lipoproteins (Vlp) of *Mycoplasma hyorhinis*, such as variants expressing longer versions of VlpA, VlpB, or VlpC are completely resistant to growth inhibition by host antibodies, unlike their shorter allelic versions [71, 72]. The same effect was observed for variable surface antigens (Vsa) of *Mycoplasma pulmonis,* where the long Vsa variants are highly resistant to complement killing while the shorter variants are susceptible [73].

From this discussion, it is not surprising that lipoproteins in *Mycoplasmataceae* have many unusual properties, including the gene lengths distribution. Being unique, these proteins cannot be assigned to any COG, which results in classification them as ORFans. Certainly, more studies should be carried out to clarify why *Mycoplasmataceae* contain long ORFans in comparison to other bacteria.

**Conclusions**

We have compared lengths' distributions of "having homologs proteins" (HHPs) and "non-having homologs proteins" (orphans or ORFans) in all currently annotated completely sequenced prokaryotic genomes.

In general, we confirmed the findings of Lipman et al. [1] established on a limited set of genomes that: (1) HHPs are, on average, longer than ORFans; (2) In a given genome, the length distribution of ORFans has a relatively narrow peak, whereas the HHPs are spread over a wider range of values. We have shown that about thirty genomes do not obey the "Lipman rules". In particular, all genomes of *Mycoplasma* and *Ureaplasma* have atypical ORFans distributions, with mean lengths of ORFan larger than the mean lengths of HHPs. We established that these differences cannot be explained by the "usual suspects" hypotheses of small genome size and low GC content of *Mycoplasmataceae. Mycoplasmataceae* are a heterogeneous group of the smallest and simplest self-replicating prokaryotes with limited metabolic capabilities, which parasitize a wide range of hosts [33, 34]. These organisms are characterized by lack of a cell wall, small genome sizes, a low GC content (23% to 40%) of the genome and usage of different genetic code (usage UGA as a tryptophan codon instead of the universal opal stop codon) [29].

We propose that the atypical features of *Mycoplasmataceae* genomes were likely developed as adaptations to their ecological niche, specifically for "quiet" co-existence with host organisms. *Mycoplasma* are known to colonize their hosts with no apparent clinical manifestations, using high variability of lipoproteins to trick the host's immune system. These are the lipoproteins that are frequently encoded by the long ORFans in *Mycoplasma* genomes, alongside with "surface protein 26-residue repeat-containing proteins" and "efflux ABC transporters". The latter functions are also associated with the obligatory parasitic lifestyle of *Mycoplasma,* which supports our hypothesis.

## Materials and Methods

### COGs database

The Clusters of Orthologous Groups of proteins (COGs) database (http://www.ncbi.nlm.nih.gov/COG/) has been a popular tool for functional annotation since its inception in 1997, particularly widely used by the microbial genomics community. The COG database is described in detail in a series of publications [74-77]. Recently, the COG-making algorithm was improved and the COG database updated [78]; however, for the purposes of our study we preferred to use the original COG repository ftp://ftp.ncbi.nlm.nih.gov/genomes/Bacteria/. This choice enabled us to compare the distributions of HHP and ORFans in as many as 1484 prokaryotic genomes, since COG functional classification of the encoded proteins is one of the required descriptors of all newly sequenced prokaryotic genomes [79].

**Statistical analysis** was conducted in R using built-in functions and custom scripts.

**List of abbreviations**

COG - a cluster of orthologous groups of genes; each COG consists of groups of proteins found to be orthologous across at least three lineages

HHP - having homologs protein; *here* COG-annotated protein

ORF – an open reading frame

ORFan – non-HHP; *here* a protein-encoding gene that is not linked to any COG

CG content - relative frequency of guanine and cytosine

$CG_3$ content - relative frequency of guanine and cytosine in the 3rd position of a codon

**Competing interests**

The authors have no competing interests

**Author's contributions**

TT and AB jointly performed data analysis, interpretation and wrote the first draft of manuscript. IL and YN helped with data interpretation and manuscript preparation.

**Reviewers' comments**

**Acknowledgements**

T.T. was supported by grants from The National Institute for General Medical Studies (GM068968), and the Eunice Kennedy Shriver National Institute of Child Health and Human Development (HD070996).

# Tables

**Table 1: Number of sequenced and annotated genomes for the selected set of bacterial species**

| Genus | Total sequenced genomes | Genomes with assigned COG |
|---|---|---|
| *Anaplasma* | 9 | 4 |
| *Ehrlichia* | 6 | 5 |
| *Neorickettsia* | 2 | 2 |
| *Wolbachia* | 7 | 4 |
| *Mycoplasma* | 68 | 29 |
| *Ureaplasma* | 3 | 3 |
| *Mesoplasma* | 2 | 1 |
| *Spiroplasma* | 6 | 0 |
| *Acholeplasma* | 3 | 1 |
| *Candidatus Phytoplasma* | 3 | 2 |
| *Anaeroplasma* | 0 | 0 |
| *Asteroplasma* | 0 | 0 |
| *Entomoplasma* | 0 | 0 |

**Table 2.** List of atypical genomes showing HHPs' average length, number of HHPs, ORFans' length, and the difference between the average length of HHP and ORFans for a given genome.

|    | SPECIES | Average length of HHPs, aa | Number of HHPs | Average length of ORFans, aa | Number ORFans | Difference between the average length of HHP and ORFans, aa |
|----|---------|---------------------------|----------------|------------------------------|---------------|------------------------------------------------------------|
| 1. | *Ureaplasma urealyticum serovar 10 ATCC 33699 uid59011* | 366 | 416 | 472 | 230 | -106 |
| 2. | *Mycoplasma genitalium G37 uid57707* | 350 | 384 | 450 | 91 | -100 |
| 3. | *Mycoplasma hyopneumoniae 7448 uid58039* | 375 | 443 | 449 | 214 | -74 |
| 4. | *Anaplasma centrale Israel uid42155* | 349 | 691 | 417 | 232 | -68 |
| 5. | *Mycoplasma gallisepticum R low uid57993* | 368 | 489 | 429 | 274 | -61 |
| 6. | *Ureaplasma parvum serovar 3 ATCC 27815 uid58887* | 359 | 413 | 410 | 196 | -51 |
| 7. | *Chlorobium chlorochromatii CaD3 uid58375* | 367 | 1564 | 417 | 435 | -50 |
| 8. | *Anaplasma marginale Florida uid58577* | 353 | 699 | 395 | 241 | -42 |
| 9. | *Mycoplasma mobile 163K uid58077* | 358 | 450 | 400 | 183 | -42 |
| 10 | *Mycoplasma hyorhinis HUB 1 uid51695* | 352 | 464 | 387 | 194 | -35 |
| 11 | *Mycoplasma hyopneumoniae 232 uid58205* | 375 | 437 | 410 | 254 | -34 |
| 12 | *Ureaplasma parvum serovar 3 ATCC 700970 uid57711* | 363 | 441 | 394 | 173 | -30 |
| 13 | *Anaplasma marginale Maries uid57629* | 352 | 699 | 382 | 249 | -30 |
| 14 | *Mycoplasma conjunctivae uid59325* | 360 | 420 | 387 | 272 | -27 |
| 15 | *Mycoplasma crocodyli MP145 uid47087* | 363 | 490 | 387 | 199 | -24 |
| 16 | *Mycoplasma hyopneumoniae J uid58059* | 391 | 471 | 413 | 186 | -23 |
| 17 | *Mycoplasma hominis ATCC 23114 uid41875* | 369 | 378 | 383 | 145 | -14 |
| 18 | *Lawsonia intracellularis PHE MN1 00 uid61575* | 492 | 51 | 500 | 53 | -8 |
| 19 | *Mycoplasma penetrans HF 2 uid57729* | 384 | 658 | 390 | 379 | -5 |
| 20 | *Burkholderia pseudomallei 1710b uid58391* | 377 | 2835 | 374 | 898 | 2 |
| 21 | *Mycoplasma putrefaciens KS1 uid72481* | 358 | 474 | 351 | 176 | 7 |
| 22 | *Mycoplasma agalactiae uid46679* | 366 | 522 | 354 | 291 | 11 |
| 23 | *Rhodobacter sphaeroides 2 4 1 uid57653* | 332 | 82 | 318 | 21 | 13 |
| 24 | *Methanobrevibacter ruminantium M1 uid45857* | 348 | 1513 | 335 | 704 | 14 |
| 25 | *Nanoarchaeum equitans Kin4 M uid58009* | 286 | 356 | 269 | 184 | 16 |
| 26 | *Mycoplasma synoviae 53 uid58061* | 363 | 479 | 345 | 180 | 17 |
| 27 | *Mycoplasma mycoides capri LC 95010 uid66189* | 384 | 619 | 361 | 303 | 23 |
| 28 | *Mycoplasma agalactiae PG2 uid61619* | 353 | 475 | 329 | 267 | 25 |
| 29 | *Mycoplasma bovis PG45 uid60859* | 371 | 526 | 343 | 239 | 28 |
| 30 | *Ehrlichia canis Jake uid58071* | 348 | 678 | 320 | 247 | 28 |
| 31 | *Mycoplasma pulmonis UAB CTIP uid61569* | 379 | 560 | 350 | 222 | 30 |
| 32 | *Mycoplasma pneumoniae M129 uid57709* | 367 | 445 | 334 | 203 | 32 |

Supplemental Table 1: Genomic GC content and genic GC$_3$ content for annotated species of *Mycoplasma, Spiroplasma*, and *Ureaplasma*.

| Organism | GC$_3$ | Genomic GC content | Genome length |
|---|---|---|---|
| *Spiroplasma apis B31 uid230613* | 0.200327 | 0.283040 | 1160554 |
| *Spiroplasma chrysopicola DF 1 uid205053* | 0.168678 | 0.287990 | 1123322 |
| *Spiroplasma diminutum CUAS 1 uid212976* | 0.113885 | 0.254591 | 945296 |
| *Spiroplasma syrphidicola EA 1 uid205054* | 0.176023 | 0.292010 | 1107344 |
| *Spiroplasma taiwanense CT 1 uid212975* | 0.137940 | 0.230113 | 11138 |
| *Spiroplasma taiwanense CT 1 uid212975* | 0.104737 | 0.238734 | 1075140 |
| *Mycoplasma agalactiae PG2 uid61619* | 0.205494 | 0.297071 | 877438 |
| *Mycoplasma agalactiae uid46679* | 0.201856 | 0.296241 | 1006702 |
| *Mycoplasma arthritidis 158L3 1 uid58005* | 0.223677 | 0.307095 | 820453 |
| *Mycoplasma bovis HB0801 uid168665* | 0.204811 | 0.293079 | 991702 |
| *Mycoplasma bovis Hubei 1 uid68691* | 0.205157 | 0.292889 | 948121 |
| *Mycoplasma bovis PG45 uid60859* | 0.202470 | 0.293147 | 1003404 |
| *Mycoplasma capricolum ATCC 27343 uid58525* | 0.092527 | 0.237704 | 1010023 |
| *Mycoplasma conjunctivae uid59325* | 0.173892 | 0.284902 | 846214 |
| *Mycoplasma crocodyli MP145 uid47087* | 0.155452 | 0.269537 | 934379 |
| *Mycoplasma cynos C142 uid184824* | 0.147378 | 0.257004 | 998123 |
| *Mycoplasma fermentans JER uid53543* | 0.154001 | 0.269460 | 977524 |
| *Mycoplasma fermentans M64 uid62099* | 0.158663 | 0.268595 | 1118751 |
| *Mycoplasma fermentans PG18 uid197154* | 0.156132 | 0.268170 | 1004014 |
| *Mycoplasma gallisepticum CA06 2006 052 5 2P uid172630* | 0.246981 | 0.316319 | 976412 |
| *Mycoplasma gallisepticum F uid162001* | 0.247519 | 0.313959 | 977612 |
| *Mycoplasma gallisepticum NC06 2006 080 5 2P uid172629* | 0.248044 | 0.316202 | 938869 |
| *Mycoplasma gallisepticum NC08 2008 031 4 3P uid172631* | 0.248610 | 0.315942 | 926650 |
| *Mycoplasma gallisepticum NC95 13295 2 2P uid172625* | 0.248706 | 0.315947 | 953989 |
| *Mycoplasma gallisepticum NC96 1596 4 2P uid172626* | 0.247410 | 0.316470 | 986257 |
| *Mycoplasma gallisepticum NY01 2001 047 5 1P uid172627* | 0.247923 | 0.316050 | 965525 |
| *Mycoplasma gallisepticum R high uid161999* | 0.242985 | 0.314674 | 1012027 |
| *Mycoplasma gallisepticum R low uid57993* | 0.243143 | 0.314699 | 1012800 |
| *Mycoplasma gallisepticum S6* | 0.246344 | 0.314669 | 985444 |
| uid200523 | | | |
| Mycoplasma gallisepticum VA94 7994 1 7P uid172624 | 0.248476 | 0.315846 | 964110 |
| Mycoplasma gallisepticum WI01 2001 043 13 2P uid172628 | 0.248994 | 0.315617 | 939844 |
| Mycoplasma genitalium G37 uid57707 | 0.235046 | 0.316891 | 580076 |
| Mycoplasma genitalium M2288 uid173372 | 0.243793 | 0.316748 | 579558 |
| Mycoplasma genitalium M2321 uid173373 | 0.243807 | 0.316742 | 579977 |
| Mycoplasma genitalium M6282 uid173371 | 0.244104 | 0.316710 | 579504 |
| Mycoplasma genitalium M6320 uid173370 | 0.244533 | 0.316767 | 579796 |
| Mycoplasma haemocanis Illinois uid82367 | 0.299288 | 0.353336 | 919992 |
| Mycoplasma haemofelis Langford 1 uid62461 | 0.354797 | 0.388514 | 1147259 |
| Mycoplasma haemofelis Ohio2 uid162029 | 0.354023 | 0.388128 | 1155937 |
| Mycoplasma hominis ATCC 23114 uid41875 | 0.165119 | 0.271182 | 665445 |
| Mycoplasma hyopneumoniae 168 L uid205052 | 0.196551 | 0.284650 | 921093 |
| Mycoplasma hyopneumoniae 168 uid162053 | 0.196624 | 0.284567 | 925576 |
| Mycoplasma hyopneumoniae 232 uid58205 | 0.196548 | 0.285612 | 892758 |
| Mycoplasma hyopneumoniae 7422 uid212968 | 0.195023 | 0.285097 | 898495 |
| Mycoplasma hyopneumoniae 7448 uid58039 | 0.194458 | 0.284889 | 920079 |
| Mycoplasma hyopneumoniae J uid58059 | 0.194191 | 0.285222 | 897405 |
| Mycoplasma hyorhinis DBS 1050 uid228933 | 0.131243 | 0.259078 | 837447 |
| Mycoplasma hyorhinis GDL 1 uid87003 | 0.130647 | 0.259062 | 837480 |
| Mycoplasma hyorhinis HUB 1 uid51695 | 0.131508 | 0.258835 | 839615 |
| Mycoplasma hyorhinis MCLD uid162087 | 0.132853 | 0.258765 | 829709 |
| Mycoplasma hyorhinis SK76 uid181997 | 0.131696 | 0.258920 | 836897 |
| Mycoplasma leachii 99 014 6 uid162031 | 0.093432 | 0.236673 | 1017232 |
| Mycoplasma leachii PG50 uid60849 | 0.092273 | 0.237505 | 1008951 |
| Mycoplasma mobile 163K uid58077 | 0.117239 | 0.249513 | 777079 |
| Mycoplasma mycoides capri LC 95010 uid66189 | 0.188954 | 0.291848 | 1840 |
| Mycoplasma mycoides capri LC 95010 uid66189 | 0.092346 | 0.238159 | 1153998 |
| Mycoplasma mycoides SC Gladysdale | 0.102919 | 0.239510 | 1193808 |
| uid197153 | | | |
| *Mycoplasma mycoides SC PG1 uid58031* | 0.098171 | 0.239656 | 1211703 |
| *Mycoplasma ovis Michigan uid232247* | 0.240392 | 0.316925 | 702511 |
| *Mycoplasma parvum Indiana uid223379* | 0.160417 | 0.269751 | 564395 |
| *Mycoplasma penetrans HF 2 uid57729* | 0.134652 | 0.257175 | 1358633 |
| *Mycoplasma pneumoniae 309 uid85495* | 0.411101 | 0.399791 | 817176 |
| *Mycoplasma pneumoniae FH uid162027* | 0.403169 | 0.400037 | 811088 |
| *Mycoplasma pneumoniae M129 B7 uid185759* | 0.408628 | 0.400088 | 816373 |
| *Mycoplasma pneumoniae M129 uid57709* | 0.411421 | 0.400080 | 816394 |
| *Mycoplasma pulmonis UAB CTIP uid61569* | 0.153015 | 0.266365 | 963879 |
| *Mycoplasma putrefaciens KS1 uid72481* | 0.139499 | 0.269392 | 832603 |
| *Mycoplasma putrefaciens Mput9231 uid198525* | 0.140889 | 0.269618 | 859996 |
| *Mycoplasma suis Illinois uid61897* | 0.199426 | 0.310762 | 742431 |
| *Mycoplasma suis KI3806 uid63665* | 0.201065 | 0.310772 | 709270 |
| *Mycoplasma synoviae 53 uid58061* | 0.166114 | 0.284952 | 799476 |
| *Mycoplasma wenyonii Massachusetts uid170731* | 0.257498 | 0.339223 | 650228 |
| *Ureaplasma parvum serovar 3 ATCC 27815 uid58887* | 0.128288 | 0.254965 | 751679 |
| *Ureaplasma parvum serovar 3 ATCC 700970 uid57711* | 0.129101 | 0.254997 | 751719 |
| *Ureaplasma urealyticum serovar 10 ATCC 33699 uid59011* | 0.126853 | 0.257742 | 874478 |